\documentclass[nohyper]{JHEP}
\usepackage{cite}

\usepackage{epsfig}

\psfull
\newskip\humongous \humongous=0pt plus 1000pt minus 1000pt
\def\caja{\mathsurround=0pt}
\def\eqalign#1{\,\vcenter{\openup1\jot
\caja   \ialign{\strut \hfil$\displaystyle{##}$&$
\displaystyle{{}##}$\hfil\crcr#1\crcr}}\,}

\newif\ifdtup

\def\eqal2#1{\,\vcenter{\openup1\jot
\caja   \ialign{\strut \hfil$\displaystyle{##}$&\hfil$
\displaystyle{{}##}$\hfil &$
\displaystyle{{}##}$\hfil\crcr#1\crcr}}\,}

\def\fun#1#2{\lower3.6pt\vbox{\baselineskip0pt\lineskip.9pt
\newcommand{\half}{{\textstyle\frac{1}{2}}}
\newcommand{\ee}{e^+e^-}
\def{\as}{\alpha_s}
\def\al{\alpha}
\newcommand{\be}{\beta}
\newcommand{\ce}{\langle C\rangle}
\def\eps{\epsilon}
\newcommand{\beq}{\begin{equation}}
\newcommand{\eeq}{\end{equation}}
\newcommand{\eq}[1]{(\ref{#1})}
\newcommand{\cl}[1]{{\cal #1}}
\newcommand{\rf}[1]{(\ref{#1})}
\newcommand{\sect}[1]{\section{#1}\setcounter{equation}{0}}
\newcommand{\secn}[1]{Section~\ref{#1}}
\newcommand{\nln}{\nonumber\\}
\newcommand{\mass}{{\mbox{\scriptsize mass}}}
\newcommand{\bub}{{\mbox{\scriptsize bub}}}
\def{\asPT}{\alpha_s^{\mbox{\tiny PT}}}
\renewcommand{\theequation}{\arabic{section}.\arabic{equation}}
\renewcommand{\labelenumi}{(\roman{enumi})}
\ialign{$\mathsurround=0pt#1\hfil##\hfil$\crcr#2\crcr\sim\crcr}}}

\def\re#1{(\ref{#1})}
\def\npp#1#2#3{{\it Nucl. \ Phys.  \ Proc. \ Suppl.}~{\bf #1} (19#3) #2}
\def\al{\alpha}
\def\be{\beta}

\def\cA{{\cal{A}}}    %   (first) coupling moment 
\def\cB{{\cal{B}}}    %   scaled B
\def\cM{{\cal{M}}}    %   Milan factor
\def\cF{{\cal{F}}}

\def\half{\mbox{\small $\frac{1}{2}$}}

\def\as{\alpha_{{\textsc{s}}}}

\def\ee{e^+e^-}

\title{Power corrections to the differential Drell-Yan cross section}

\author{Mrinal Dasgupta \\
  Dipartimento di Fisica, Universit\`a di Milano Bicocca \\
  and INFN, Sezione di Milano,\\ 
  Via Celoria 16, I-20133, Milano, Italy \\
  E-mail: \email{dasgupta@mi.infn.it}}

\abstract{We estimate the power corrections (infrared renormalon contributions)
to the coefficient functions for the differential Drell-Yan
cross-section $d^2\sigma/dQ^2dy$, where $Q^2$ is the mass squared and
$y$ the rapidity of the produced lepton pair. We employ the dispersive 
method based on the analysis of one-loop Feynman graphs
containing a massive gluon.}

\preprint{Bicocca--FT--99--33\\
          November 1999}

\begin{document}

\section{Introduction}
\label{intro}
 The Drell-Yan process \cite{DY} describes the
collision of two hadrons and the subsequent production of a lepton
pair and a hadronic final state. In the parton model it proceeds
simply via the annihilation mechanism where a quark and anti-quark
generated by the parent hadrons annihilate to form a photon which
decays to a lepton pair.  At present day collider energies one is also
able to produce the electroweak W and Z bosons on mass shell via this
mechanism.

Historically the Drell-Yan process has played an important role in the
development of QCD. The parton model became more firmly established
when it was realised that it gave a good description of the data in
hadron-hadron collisions. In addition it became evident that QCD
perturbation theory could be applied to describe strong interaction
phenomena when one encountered the very same mass singularities in
Drell-Yan calculations as those in the case of deeply inelastic
scattering giving rise to the concept of universal functions that
control the long-distance dynamics, which one refers to as parton
distributions.

Data from hadron-hadron collisions has proved a valuable source from
which one has been able to constrain and measure various parton
distributions. In particular such data provides a means to extract
information on the quark distributions in pions which is inaccessible
from DIS experiments. Data on low-mass lepton pair production has been
used to study the small $x$ behaviour of parton distributions. For a
thorough review of available Drell-Yan data and comparisons with
theory the reader is referred to Ref.~\cite{JamWal}.

In this article we choose to concentrate our attention on the cross
section $d^2 \sigma/d Q^2 dy $ with $Q^2$ being the mass squared and
$y$ the rapidity of the produced lepton pair. Most recently the CDF
collaboration have studied the rapidity dependence of the Drell-Yan
cross section in a limited rapidity range \cite{Abe}.  In fact from
the experimental side there is a wealth of data on various rapidity
distributions (see Ref.~\cite{JamWal} ) but progress on the
theoretical side is somewhat lacking. While the ${\mathcal{O}}(\alpha_s)$
perturbative QCD calculations for the above distribution were
performed several years ago \cite{Ellis,French} there is as yet no
${\mathcal{O}}(\alpha_s^2)$ estimate. Also lacking is any estimate of the
non-perturbative power-like corrections which have been extensively
studied in many other cases . The aim of this article is to study the
power correction to the rapidity distribution which
should in principle allow a better description of the data when added
to the perturbative predictions.  Power correction predictions already
exist for the more inclusive cross section $d \sigma/dQ^2$ where a $1/Q^2$
dependence is predicted with a characteristic phase space enhancement
\cite{BPY,BenBrau} .
 
 From a purely theoretical viewpoint the motivation of the work
 described here is the testing of current ideas on power corrections
 which seem to indicate that though these contributions are
 non-perturbative in origin one may suitably extend a perturbative
 approach to estimate them. The success of such an approach in the
 case of DIS structure functions \cite{dasweb96} is encouraging
 enough to extend this study to the present case. In view of the
 relative simplicity of the renormalon calculations that yield
 predictions for the power corrections and the range and accuracy of
 current experimental data on several different QCD observables one
 should be able to undertake a serious and extensive confrontation of
 these theoretical ideas with the data, a task that has already begun
 \cite{BY,shapee,shapdis}.

This paper is organised as follows. In the next section we give a very
brief review of the dispersive treatment of power corrections which
has been described in great detail previously \cite{BPY}. Following
this we mention the notation and introduce the kinematical variables
relevant to our study. In the following section we describe our
results for the power corrections and finally make some concluding
remarks.

\section{Dispersive Approach}
\label{dispv}
 The main ideas of the dispersive
approach to power corrections can be briefly summed up as below.
First one assumes a QED inspired dispersion relation to be formally
true in the QCD case so that
\begin{equation}
\label{disp}
\alpha_s(k^2) = - \int_{0}^{\infty}\frac{d \mu^2}{\mu^2+k^2} \rho_s(\mu^2)
\end{equation}
with the spectral function
\begin{equation}
\label{spec}
\rho_s(\mu^2)= \frac{1}{2 \pi i} \left \{ \alpha_s(\mu^2 e^{i \pi})-\alpha_s(\mu^2 \;e^{-i \pi})\right \}\; .
\end{equation}
Thus one is assuming that the QCD coupling is well behaved in the infra-red and the only singularity is a discontinuity on the negative real axis of its argument. 

Non-perturbative effects at long distances are expected to give rise to a modification to the perturbatively-calculated strong coupling at low scales, $\delta\alpha_s(\mu^2) = \alpha_s(\mu^2) - \alpha_s^{\mbox{\tiny{PT}}}(\mu^2)$, $\alpha_s^{\mbox{\tiny{PT}}}(\mu^2)$ being the perturbatively-calculated running coupling \cite{BPY}. 
The spectral function \re{spec} receives the corresponding modification 
\begin{equation}
\label{mod}
\delta \rho_s(\mu^2) = \frac{1}{2 \pi i} {\mbox{Disc}} \left \{ \delta \alpha_s (-\mu^2) \right \} \;.
\end{equation}
To consider the effect of the above on an observable $F$ one assumes the implicit inclusion of a gauge invariant set of higher order graphs (which in a large $N_f$ approximation just reduce to quark bubble insertions ) combined with single gluon emission has the effect of generating the running coupling in the one-loop calculation of $F$ . In practice the running coupling can only be reconstructed in this manner, for a sufficiently inclusive observable like the one we study in this article. The above considerations allow one to write \cite{BPY}
\begin{equation}
F = \int_0^{\infty} \frac{d \mu^2}{\mu^2} \rho_s(\mu^2) \left [ {\mathcal{F}}\left ( \frac{\mu^2}{Q^2} \right ) - {\mathcal{F}} (0) \right ]
\end{equation}
where ${\mathcal{F}}$  is the one loop correction to the observable, computed with a finite gluon mass $\mu^2$. Then the non-perturbative region contributes to $F$ through Eq.~\re{mod} and one finds 
\begin{equation}
\label{disc}
 \delta{F} = \int_{0}^{\infty} \frac{d \mu^2}{\mu^2} \delta \alpha_s (\mu^2) {\mathcal{G}} (\mu^2/Q^2)
\end{equation}
where setting $\mu^2/Q^2 = \epsilon $
\begin{equation}
\label{discone}
{\mathcal{G}(\epsilon)} = -\frac{1}{2 \pi i} {\mbox{Disc}} \left \{ {\mathcal{F}}(-\epsilon)\right \}\;.
\end{equation}
Since $\delta \alpha_s (\mu^2)$ is limited to low values of $\mu^2$ the asymptotic behaviour of $\delta F$ at high $Q^2$ is given by its behaviour in the limit $\epsilon \rightarrow 0$. Clearly only terms that are non-analytic in $\epsilon$ in the small $\epsilon$ behaviour of ${\mathcal{F}}$, yield non-perturbative modifications to $F$ within the above approach.
In the present case the characteristic function $\cF$ will be found to have the general small $\epsilon$ behaviour 
\begin{equation}
\cF \sim \frac{C_F}{2\pi} \left ( C_1\left \{ x \right \} \epsilon \ln^2\epsilon + C_2 \left \{x \right \} \epsilon \ln \epsilon \right )
\end{equation}
with $\left \{ x \right \}$ denoting the phase space dependence that we compute here.
According to \re{disc} and \re{discone} the above $\epsilon$ dependence translates into the power correction
\begin{equation}
\frac{\cA_2}{Q^2} \left [ C_2 \left \{ x \right \}+2 C_1 \left \{x \right \}\ln \left (\frac{\cB_2}{Q^2} \right ) \right ]
\end{equation}
with the parameters $\cA_2$ and $\cB_2$ being 
\begin{eqnarray}
\cA_2 &=& \frac{C_F}{2\pi}\int_0^\infty \frac{d\mu^2}{\mu^2}\,\mu^2\,\delta\alpha_s(\mu^2) \\ \nonumber
\ln\cB_2 &=& \frac{1}{\cA_2}\frac{C_F}{2\pi}\int_0^\infty \frac{d\mu^2}{\mu^2}\,\mu^2\,\ln\mu^2 \; \delta\alpha_s(\mu^2) \; .
\end{eqnarray}
Then one invokes the universality assumption, in that if one believes that the concept of $\alpha_s$ can be meaningfully extended to small scales, it can be done in an observable and indeed process independent fashion. 
This would allow us to extract the value of the above defined moments of the coupling, $\cA_2$ and $\cB_2$, from any experimental data where the same power correction is obtained and use it to fit Drell-Yan data. 
Recent studies of $1/Q$ corrections to event shape variables in
$e^{+}e^{-}$ annihilation \cite{shapee} and in DIS \cite{shapdis} lend support to this idea .
Studies of the $1/Q^2$ non-singlet contribution to DIS structure functions suggest that $\cA_2 \simeq 0.2 \; \mbox{GeV}^2$ \cite{dasweb96}. We do not know the value of the parameter $\cB_2$ from any experimental data and hence it can be treated as a free parameter in the fit to the Drell-Yan case.

On the other hand as has been explained in detail in previous articles (see for instance Ref.~ \cite{web}) the parameter $\cA_2$ and indeed any other moment of the modified coupling, $\delta \alpha_s$, depends on the order of the perturbative result with which one is interested in merging the non-perturbative piece. The above DIS value is relevant to merging with the next-to--leading  order (NLO) QCD prediction while in the present case of the differential Drell-Yan cross-section one only knows the leading QCD correction to the parton model. However we use the DIS value for illustrative purposes here and in the expectation that the NLO result will be available soon.

\section{Definitions and Kinematics}
We wish to compute the power corrections to the observable
$d^2\sigma/dQ^2 dy$ where $Q^2$ is the mass squared and $y$ is
the rapidity of the produced lepton pair. To lowest order the
expression for the above quantity is simply 
 
\begin{equation} 
\frac{d^2 \sigma}{dQ^2 dy} = \frac{\sigma_0}{Ns} \left [ \sum_q e_q^2
  \left \{f_{q/A} (z_1,Q^2)f_{\bar{q}/B}(z_2,Q^2) + (q \leftrightarrow
\bar{q} )\right \} \right ] \;, 
\end{equation}
where $$\sigma_0 = \frac{4 \pi \alpha^2}{3 Q^2},$$ $N$ is the
number of colours and $s$ the center-of--mass energy squared. In addition $e_q$ refers to the charge of flavour
$q$ in the flavour sum above and $z_1,z_2$ refer to
the momentum fractions carried by the quark and antiquark, of the
parent hadrons $A$ and $B$ while the $f$ functions are the
corresponding parton distributions. We explicitly have, in the hadronic
centre of mass frame, the parton momenta (neglecting their small intrinsic $k_t$)
\begin{equation} \eqalign{
&p_1 = \frac{\sqrt{s}}{2}(z_1,0,0,z_1) \cr&
p_2 = \frac{\sqrt{s}}{2}(z_2,0,0,-z_2)}
\end{equation}
The invariant mass of the produced lepton pair is then simply given by 
\begin{equation}
Q^2 = (p_1+p_2)^2 = s z_1 z_2
\end{equation}
while its rapidity is 
\begin{equation}
y = \frac{1}{2} \ln\left (\frac{E+P_L}{E-P_L} \right )= \frac{1}{2} \ln \left ( \frac{z_1}{z_2} \right )
\end{equation}
which yield
\begin{equation}
\label{taux}
z_1 =\sqrt{\tau}e^{y} \:,\: z_2 = \sqrt{\tau} e^{-y} \:,\: \tau = \frac{Q^2}{s}.
\end{equation}
Beyond the naive parton model one has to consider radiative corrections to the above picture, wherein initial partons carrying momentum fractions $z_1$ and $z_2$ degrade their momenta via gluon radiation before annihilating to form the lepton pair. Hence the simple correspondence \re{taux} between the momentum fractions of the initiating partons and the mass and rapidity of the lepton pair no longer holds beyond lowest order. We define for later use kinematic
variables $x_1$ and $x_2$ such that 
\begin{equation} 
x_1 =\sqrt{\tau} e^{y} \;, \;  x_2 = \sqrt{\tau}e^{-y}
\end{equation}
and which at Born level are just equal to the initial parton momentum fractions.

For computing the renormalon contribution to the desired observable we
shall need the squared matrix element for the QCD radiative process
\footnote{For computing the power correction we are only considering
  the annihilation contribution. In general there is also the QCD Compton
  scattering process to consider, beyond the naive parton model. We shall
  comment further on this in the final section.}
\begin{equation}
q(p_1)+\bar{q}(p_2) \rightarrow \gamma^{*}(q)+g(k),  
\end{equation}
which takes the simple form
\begin{equation}
\label{ME}
{\mathcal{M}}_{DY} = \frac{\hat{s}^2+Q^4(1+\epsilon)^2}{\hat{u}\hat{t}}-2-\epsilon Q^4 \left (\frac{1}{\hat{u}^2}+\frac{1}{\hat{t}^2} \right )
\end{equation}
where $\hat{s},\hat{t},\hat{u}$ are Mandelstam invariants defined as
\begin{equation}
\hat{s} = (p_1+p_2)^2,\; \hat{t} = (p_1-k)^2, \; \hat{u} = (p_1-q)^2 ,
\end{equation}
and which satisfy the relation
\begin{equation}
\hat{s}+\hat{t}+\hat{u} = Q^2 + \mu^2
\end{equation}
with $\mu^2=\epsilon Q^2$ being the squared gluon mass and $ \hat{s}=z_1 z_2 s $.

The rapidity $y$ can readily be obtained in terms of the Mandelstam
invariants and the momentum fractions $z_1$ and $z_2$ as
\begin{equation}
\label{rap}
 y = \frac{1}{2} \ln \left (\frac{Q^2-\hat{t}}{\hat{s}+\hat{t}-\mu^2} \; \frac{z_1}{z_2} \right )
\end{equation}   
which is equivalent to 
\begin{eqnarray}
\label{cat}
\hat{t} &=& \frac{Q^2 x_2 z_1 -x_1 z_2(\hat{s}-\mu^2)}{x_2 z_1+ x_1 z_2}\\
\hat{u} &=& \frac{Q^2 x_1 z_2 -x_2 z_1(\hat{s}-\mu^2)}{x_2 z_1 + x_1 z_2} \nonumber
\end{eqnarray}
with $x_1$ and $x_2$ defined as before. 
The phase space for massive gluon emission is given by
\begin{equation}
(z_1-x_1)(z_2-x_2) \geq \frac{\mu^2}{s}.
\end{equation}

Putting the above together we obtain the required differential 
cross section
\begin{equation}
\frac{d^2 \hat{\sigma}}{dQ^2 dy} = A \frac{ \tau( \tau+z_1 z_2-\epsilon \tau )}{(z_1 z_2) (z_1 x_2 +z_2 x_1)^2} \; {\mathcal{M}}(s,x_1,x_2,z_1,z_2,\epsilon)
 \end{equation}
where $\cM$ is just the matrix element \re{ME} 
with the substitutions
of Eq.~(3.12) and $ A = 16 \alpha^2 \alpha_s e_q^2/27Q^2 s$.
The hadron level result is related to the corresponding 
partonic quantity by folding 
with parton distribution functions as below
\begin{equation}
\label{fac}
\frac{d^2 \sigma}{d Q^2 dy} = \sum_q \int_{x_1}^{1}\int_{x_2}^{1} dz_1 dz_2 \frac{d^2 \hat{\sigma}}{dQ^2 dy} \Theta \left ( (z_1-x_1)(z_2-x_2)-\frac{\mu^2}{s}\right ) {\mathcal{F}}_{q}(z_1,z_2)
\end{equation}
where for brevity we defined 
\begin{equation}
{\mathcal{F}}_q(z_1,z_2) = \left \{f_{q/A}(z_1,Q^2)f_{\bar{q}/B}(z_2,Q^2) + (q \leftrightarrow
\bar{q} )\right \} \;. 
\end{equation}

Next we introduce the variables $\xi = x_1/z_1 $ and $\zeta = x_2/z_2 $ in terms of which Eq.~\re{fac} assumes the familiar form  
\begin{equation}
\label{conv}
\frac{d^2 \sigma}{dQ^2 dy} = \sum_q e_q^2 \frac{C_F \alpha_s}{2 \pi}\int_{x_1}^{1}\int_{x_2}^{1} \frac{d \xi}{\xi} \frac{d \zeta}{\zeta} C(\xi,\zeta,\epsilon){\mathcal{F}}(x_1/\xi,x_2/\zeta)\Theta((1-\xi)(1-\zeta)-\epsilon \xi \zeta)
\end{equation} 
in writing which we have dropped an uninteresting overall constant
factor
$\sigma_0/Ns$, which also appears in the Born cross section to
which we shall normalise the result subsequently. The coefficient
function $C(\xi,\zeta,\epsilon)$ then takes the form 
\begin{eqnarray}
\label{long}
C(\xi,\zeta,\epsilon) &=& \frac{2(1+ \xi \zeta -\epsilon \xi \zeta)(1+\xi^2 \zeta^2(1+\epsilon)^2)}{(1-\xi^2-\epsilon \xi \zeta)(1-\zeta^2 -\epsilon \xi \zeta)}-4  \frac{\xi \zeta (1+\xi \zeta-\epsilon \xi \zeta)}{(\xi+\zeta)^2}\\ 
&-& 2\epsilon \xi \zeta (1 + \xi \zeta -\epsilon \xi \zeta) \left [ \frac{\xi^2}{(\xi^2-1+\epsilon\xi\zeta)^2} +\frac{\zeta^2}{(\zeta^2-1+\epsilon \xi \zeta)^2}\right ]\nonumber.
\end{eqnarray}
 
Next taking moments of the convolution equation  \re{conv} we find as
usual (note the different powers)
\begin{equation}
\label{mom}
\int_{0}^{1}\int_{0}^{1} dx_1 \; dx_2 \; x_1^{N} \; x_2^{M} \frac{d^2 \sigma}{dQ^2 dy}= \sum_q e_q^2
\frac{C_F \alpha_s}{2 \pi} \;\tilde{C}(N,M,\epsilon) \; \tilde{{\mathcal{F}}}_{q}(N,M)
\end{equation} 
where 
\begin{eqnarray}
\tilde{C}(N,M,\epsilon) &=& \int_{0}^{1}\int_{0}^{1} \xi^N \zeta^M \; C(\xi,\zeta,\epsilon) \; \Theta((1-\xi)(1-\zeta)-\epsilon \xi \zeta) d\xi \;d\zeta \\
\tilde{{\mathcal{F}}}_q (N,M) &=&  \tilde{f}_{q/A}(N) \; \tilde{f}_{\bar{q}/B} (M) + (q \leftrightarrow \bar{q})  \nonumber
\end{eqnarray}
and the $\tilde{f}$ functions represent the Mellin transforms of the quark and anti-quark density functions.

As explained previously, to extract the power corrections we have to look for non-analytic behaviour in $\epsilon$, in the small $\epsilon$ expansion of the Mellin transformed coefficient functions $\tilde{C}(N,M,\epsilon)$. This non-analyticity will manifest itself (in the present case ) through the appearance of a logarithmic dependence on $\epsilon$ in addition to the usual logarithmic divergences generated by the soft and collinear regions of integration. These divergences are of course cancelled by virtual corrections in the case of the infrared and by absorbing the collinear divergences into the definition of the parton densities. We are then left with terms like $\epsilon \ln^2 \epsilon$ and $\epsilon \ln \epsilon$ which are in one to one correspondence with infrared renormalon poles in the Borel plane and  will generate $1/Q^2$ power corrections. We neglect similar terms that appear at order $\epsilon^2$ and higher orders in $\epsilon$ since they will induce sub-leading ${\mathcal{O}}(1/Q^4)$ power corrections in which we are not phenomenologically interested.

\section{Power Corrections}
In taking the Mellin transforms of the coefficient functions as
mentioned in \re{mom} one finds that the second and third pieces of the
expression on the right-hand--side of Eq.~\re{long} do not produce any
logarithmic divergences but contribute only to the power corrections
through the appearance of an $\epsilon \ln \epsilon $  term. The
collinear and infrared divergences lie in the first piece on the
right-hand--side  of \re{long}. The Mellin transforms of this piece
are cumbersome to evaluate directly and the result is most easily arrived at after a further change of integration variables: 
\begin{equation}
\hat{\tau} = \xi \zeta, \: {\eta} = \frac{\xi}{\zeta}.
\end{equation}

In terms of these variables the double Mellin transform of the above mentioned piece takes the simple form 
\begin{equation}
\int_{0}^{1/(1+\sqrt{\epsilon})^2} d\hat{\tau}\int_{\eta_1}^{\eta_2} d{{\eta}} \; \hat{\tau}^{(N+M)/2}\; \eta^{(N-M)/2}\;\frac{(1+\hat{\tau}-\epsilon \hat{\tau})(1+\hat{\tau}^2(1+\epsilon)^2)}{(1-\eta \hat{\tau}-\epsilon \hat{\tau})(\eta - \hat{\tau}-\epsilon\eta\hat{\tau})}.
\end{equation}
In the above equation the limits of the $\eta$ integration are given by
\begin{equation}
\label{ph}
\eta_1 = \frac{1}{\eta_2}=\frac{1}{2 \hat{\tau}}
(\Lambda+2\hat{\tau}+\sqrt{\Lambda^2+4\Lambda \hat{\tau}}),\: \:
\Lambda = 1-2\hat{\tau}+\hat{\tau}^2-2\hat{\tau}\epsilon -2 \hat{\tau}^2 \epsilon +\epsilon^2 \hat{\tau}^2.
\end{equation}

Taking the $\eta$ moments is now an easy task but inserting the limits \re{ph} complicates the extraction of the $\tau$ moments. In particular one has to evaluate the form (apart from other relatively straightforward integrals)  
\begin{equation}
\int_{0}^{1/(1+\sqrt{\epsilon})^2}d\hat{\tau}\;\hat{\tau}^{\omega}\left (\frac{1+\hat{\tau}^2+\hat{\tau}^2 \epsilon^2 +2 \epsilon \hat{\tau}^2}{1-\hat{\tau}-\epsilon \hat{\tau}} \right )\; \tanh^{-1} \left [\frac{\sqrt{\Lambda}}{1-\hat{\tau}-\epsilon \hat{\tau}} \right ]
\end{equation}
where we use $\omega$ to denote a generic power which  depends on $M$ and $N$.
The above form has been evaluated in Ref.~\cite{BPY} and we shall not
describe the manipulations that lead to the evaluation of integrals of
the above type but refer the reader to section 4.6 and appendix A of Ref.~\cite{BPY} for the details.   
The moments of the second and third pieces on the right hand side of Eq.~\re{long} are most easily evaluated directly in $\xi,\zeta$ space. 

Putting together the contribution of all the pieces and including the
time-like virtual corrections (which are independent of $M$ and $N$
and were computed in Ref.~\cite{BPY})
we obtain the result for the total contribution as follows:
\begin{equation}
\label{master}
\tilde{C}^{{\mathrm{R+V}}}(N,M,\epsilon) = \tilde{C}_0\;\ln{\epsilon}+\tilde{C}_1 \; \epsilon\ln^2{\epsilon}+\tilde{C}_2\epsilon \ln{\epsilon}
\end{equation}
where
\begin{eqnarray}
\label{vlong}
\tilde{C}_0 &=& \left [ 
  2S_1(N+1)+2S_1(M+1)-3+\frac{1}{M+1}+\frac{1}{N+1}+\frac{1}{N+2}+\frac{1}{M+2} \right ] \\ 
\tilde{C}_1 &=& - \left [ \frac{N}{2} +\frac{M}{2} +3 \right ]\\ 
\tilde{C}_2 &=& \left [ -(M+N+2)S_1(N+1)-(M+N+2)S_1(M+1)+2MN+5M+5N+4
 \right. \nonumber \\ 
            & & +\frac{N-M}{N+1}+\frac{M-N}{M+1}+\frac{N-2M}{2(N+2)}+\frac{M-2N}{2(M+2)} \\
            & & +\left. \frac{N-2M-3}{2(N+3)}+\frac{M-2N-3}{2(M+3)}
            \right ] \nonumber
\end{eqnarray}
and
\begin{equation}
S_1(N) = \sum_{j=1}^{N-1}\frac{1}{j} = \psi(N)+\gamma_E.
\end{equation}

Notice the absence of a $\ln^2 \epsilon$ term which cancels in the sum 
of the real and virtual pieces.  In addition, the logarithmic divergence in the cut-off (gluon mass) is absorbed into the structure functions and will not concern us any more. We shall instead concentrate our attention on the corrections generated by the $\epsilon \ln^2{\epsilon}$ and $\epsilon \ln{\epsilon}$ piece of Eq.~\re{master}.  
As a check on our result above, we note that in the diagonal limit
$M=N$ we recover the result obtained in Ref.~\cite{BPY} provided one
changes $N$ to $N-1$ in accordance with the different definition of
the moments adopted in that reference. This agreement is expected as in the special case, $M=N$, our calculation reduces to just extracting the $\tau$ dependence of the power corrections.
 The  above results can be easily expressed in $\xi,\zeta$ space and the convolutions with the parton densities performed as prescribed in Eq.~\re{conv}. 

\section{Discussion}
 For the purposes of illustration we assume that $f_{q/A} =
 f_{\bar{q}/B} =q$ with the $q \leftrightarrow \bar{q}$ term being
 similarly labelled $\bar{q}$. In reality of course such assumptions
 about the density functions will depend on the beam and the target
 used in the relevant experiment but our predictions can be easily
 adjusted to every case.   

\FIGURE{\epsfig{file=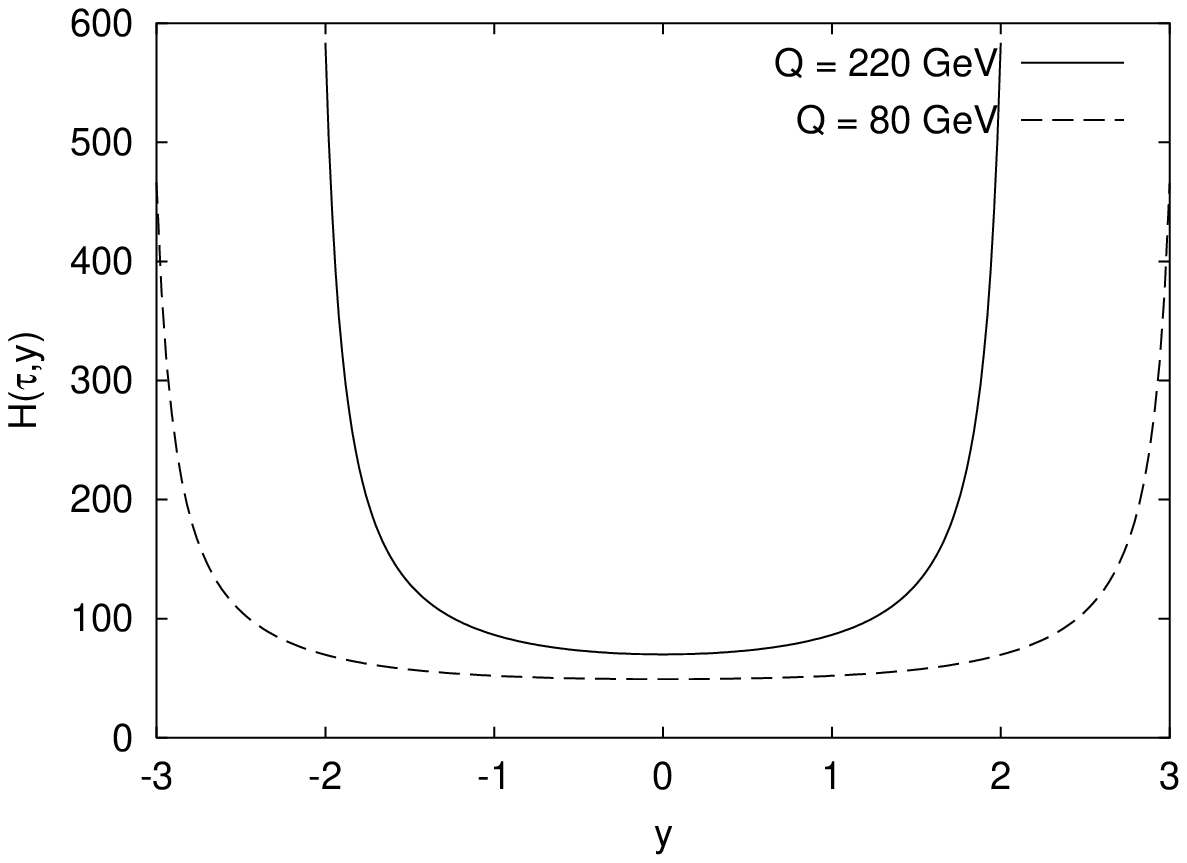}\\
  \vspace{0.05cm}
  \label{fig:kin}
  \caption{Plot of the rapidity dependence of the power
    correction function $H(\tau,y)$. The solid line corresponds to
    $\tau=0.0149,\; |y| < 2 $ while the dashed line corresponds to $\tau
    = 0.0019,\; |y| < 3$.}}
   
Then we find that performing the convolution and extracting the lowest (Born) 
order result allows one to write :
\begin{equation}
\frac{d^2 \sigma}{dQ^2 dy} = \left ( \frac{d^2 \sigma}{dQ^2 dy} \right )_{B} \left [1+\frac{\cA_2}{Q^2} \; H(x_1,x_2) \right ]
\end{equation} 
where the power correction function $H(x_1,x_2)$ can be expressed as
\begin{equation}
\label{trial}
\eqalign{H(x_1,x_2) =\frac{1}{f(x_1,x_2)}
\sum_{q} e_q^2 [ &\left \{x_1 \;q^{\prime}(x_1)q(x_2)+x_2 \; q^{\prime}(x_2)q(x_1)-4 \; q(x_1)q(x_2)\right \}\ln \left ( \frac{\cB_2}{Q^2} \right )\cr
                 &+ 2x_1x_2\; q^{\prime}(x_1)q^{\prime}(x_2)-3x_1\;q^{\prime}(x_1)q(x_2)-3 x_2 \;q^{\prime}(x_2)q(x_1) \cr 
&-q(x_1) \; I(x_2) -q(x_2) \;I(x_1) -x_1 
\;q^{\prime}(x_1)\;K(x_2) \cr     
&-x_2 \; q^{\prime}(x_2)\; K(x_1) +q \leftrightarrow \bar{q} ] }
\end{equation}
with the following definitions:
\begin{eqnarray}
f(x_1,x_2) &=& \sum_{q}e_q^2 (q(x_1)q(x_2)+q \leftrightarrow \bar{q}) \\ 
I(z)&=& \int_{z}^{1} \frac{dy}{y} \left [ \frac{1}{(1-y)_+} \left (\frac{z}{y} q^{\prime} \left (\frac{z}{y} \right )\right )
+2y^2 \; q \left (\frac{z}{y} \right )\right ] \\
K(z) &=& \int_{z}^{1} \frac{dy}{y} \left [ \frac{1}{(1-y)}_+ -(1+y+y^2) \right ]q \left( \frac{z}{y} \right ).
\end{eqnarray}

In the above formulae the `+' distributions have their usual meaning, 
the prime symbol on the $q$ denotes a derivative and the scale
($Q^2$) dependence of the parton distributions is understood.  Note that the
symmetry under the exchange of $x_1$ and $x_2$ implies that the power
correction is a symmetric function about $y=0$ with our assumption
about the parton densities, which would be valid in the $p\bar{p}$
case. For a general beam and target, asymmetry will be induced purely
by the differing parton densities in the beam and target.

A plot of the function $H$  against the rapidity $y$ is shown in figure 1, for two different $Q$ values which correspond to two 
different values of $\tau$. 
The value of $\sqrt{s}$ was chosen to be $1.8 \; {\mbox{TeV}}$, and the
plots were made using the MRSA valence parton distributions \cite{MRS}.
The value of the parameter $\cB_2$ was chosen to be $1 \;
{\mbox{GeV}}^2$ for illustrative purposes; in principle 
 there is no reason why
it should have a value close to the one chosen here. 

Both curves shown reflect a similar
behaviour, namely that the power correction is quite flat until one
starts to reach the edge of the rapidity range shown in either
case. For example, in the plot at $80 \; {\mbox{GeV}}$ as one gets closer to $y = 3$, 
we start approaching the region where $x_1 = \sqrt{\tau} e^y$ is near
unity. In fact for $y = 2.9$ and $Q = 80 \; {\mbox{GeV}}$ one finds $x_1
  =0.80$. Similarly as one progresses towards more negative rapidity
  values, 
  $x_2 =
  \sqrt{\tau}e^{-y}$ starts approaching unity.
When either $x_1$ or $x_2$ gets close to unity an explosive behaviour of 
the power correction is witnessed. This behaviour is a
reflection of the singular nature (as $\xi,\zeta \rightarrow 1$) of
the derivatives of delta functions, which are obtained on inverse
transforming the Mellin-space coefficient functions $\tilde{C}_1$ and $\tilde{C}_2$. 

At the edge of the rapidity range for the $80 \; {\mbox{GeV}}$ plot, the
overall effect of the power correction $\cA_2 H(\tau,y)/Q^2$ is large enough to be comparable  
to ${\alpha_s^2(Q)}$ which makes it an important effect to consider
while comparing perturbative predictions with the data. For larger $Q$ values, although there is a logarithmic 
enhancement of the power correction in this case, the suppression by
$Q^2$ should reduce the significance of the correction. 

In the more
inclusive case of the cross-section differential in $Q^2$ the
explosion of the power correction will only be important as
{\it{both}} 
$x_1$ 
and $x_2$ approach unity, in other words in the limit $\tau
\rightarrow 1$ \cite{BPY}. This region is probably beyond
any experimental interest as the $Q$ value is too high (close to the
centre of mass energy) in that case.
Hence while the
power correction should not be an important consideration in the $Q^2$
distribution, its presence should be felt in the combined $Q^2,y$ 
distribution, especially at moderate $Q$ values and towards the edge
of the allowed rapidity range.

Lastly we comment on the fact that at higher rapidities one would
expect the QCD Compton scattering process to become significant. We have
not taken this into account here as it is not yet completely clear how to treat
renormalon contributions from incoming gluons as a genuinely gluonic
contribution. The only treatment suggested till now is to compute the
renormalon contribution, treating the gluon as an internal line
radiated off an incoming quark which has been done for 
singlet DIS contributions (see Ref.~\cite{smy} and references
therein). This procedure can in principle be applied here in the
Drell-Yan case.
However it leaves the question of how one may unambiguously factor off the quark to gluon
splitting in order to be able to convolute with the gluon density,
which rapidly grows at small momentum fractions $x$. Till this issue
is better understood we postpone further discussion on this topic.
\medskip
\medskip
\medskip

\noindent{\large\bf Acknowledgements}
I would like to thank Bryan Webber, Giuseppe Marchesini and Gavin
Salam for many useful discussions.
This work was supported in part by the EU Fourth Framework Programme 
`Training and Mobility of Researchers', Network `Quantum Chromodynamics and 
the Deep Structure of Elementary Particles', contract FMRX-CT98-0194 
(DG 12 - MIHT).

\end{document}